\documentclass[prl,preprint,amsmath,amssymb, aps]{revtex4-1}

\usepackage{graphicx}
\usepackage[space]{grffile}
\usepackage{latexsym}
\usepackage{amsfonts,amsmath,amssymb}
\usepackage{url}
\usepackage[utf8]{inputenc}
\usepackage{fancyref}
\usepackage{hyperref}
\hypersetup{colorlinks=false,pdfborder={0 0 0},}
\usepackage{textcomp}
\usepackage{longtable}
\usepackage{multirow,booktabs}

\usepackage{natbib}

\begin{document}

\title{Top-quark electroweak couplings at the FCC-ee}

\author{{\textbf{Patrick Janot}}\\{\emph{CERN, PH Department, Geneva,
Switzerland}}}
\affiliation{}

\bibliographystyle{apsrev4-1}

\begin{abstract}
An optimal-observable analysis of the lepton angular and energy
distributions from top-quark pair production with semi-leptonic decays
in ${\rm e^+e^-}$ collisions is used to predict the potential
sensitivity of the FCC-ee to the couplings of the top quark to the
photon and the Z.

\end{abstract}

\maketitle

\section{Introduction}
\label{sec:Introduction}

The design study of the Future Circular Colliders (FCC) in a 100-km ring in the Geneva area has started at CERN at the beginning of 2014, as an option for post-LHC particle accelerators. The study has an emphasis on proton-proton and electron-positron high-energy frontier machines~\cite{FCCWebSite}. In the current plans, the first step of the FCC physics programme would exploit a high-luminosity ${\rm e^+e^-}$ collider called FCC-ee, with centre-of-mass energies ranging from below the Z pole to the ${\rm t\bar t}$ threshold and beyond. A first look at the physics case of the FCC-ee can be found in Ref.~\cite{Bicer_2014}.

In this first look, the focus regarding top-quark physics was on precision measurements of the top-quark mass, width, and Yukawa coupling through a scan of the ${\rm t\bar t}$ production threshold, with $\sqrt{s}$ comprised between~340 and~350\,GeV. The expected precision on the top-quark mass was in turn used, together with the outstanding precisions on the Z peak observables and on the W mass, in a global electroweak fit to set constraints on weakly-coupled new physics up to a scale of 100\,TeV. Although not studied in the first look, measurements of the top-quark electroweak couplings are of interest, as new physics might also show up via significant deviations of these couplings with respect to their standard-model predictions. Theories in which the top quark and the Higgs boson are composite lead to such deviations. The inclusion of a direct measurement of the ttZ coupling in the global electroweak fit is therefore likely to further constrain these theories. 

It has been claimed that both a centre-of-mass energy well beyond the top-quark pair production threshold and a large longitudinal polarization of the incoming electron and positron beams are crucially needed to independently access the tt$\gamma$ and the ttZ couplings for both chirality states of the top quark. In Ref.~\cite{Baer_2013}, it is shown that the measurements of the total event rate and the forward-backward asymmetry of the top quark, with 500\,${\rm fb}^{-1}$ at $\sqrt{s}=500$\,GeV and with beam polarizations of ${\cal P} = \pm 0.8$, ${\cal P}^\prime = \mp 0.3$, allow for this distinction. 

The aforementioned claim is revisited in the present study. The sensitivity to the top-quark electroweak couplings is estimated here with an optimal-observable analysis of the lepton angular and energy distributions of over a million events from ${\rm t\bar t}$ production at the FCC-ee, in the $\ell \nu {\rm q \bar q b \bar b}$ final states (with $\ell = {\rm e}$ or $\mu$), without incoming beam polarization and with a centre-of-mass energy not significantly above the  ${\rm t\bar t}$ production threshold. 

Such a sensitivity can be understood from the fact that the top-quark polarization arising from its coupling to the Z is maximally transferred to the final state particles via the weak top-quark decay ${\rm t \to W b}$ with a 100\% branching fraction: the lack of initial polarization is compensated by the presence of substantial final state polarization, and by a larger integrated luminosity. A similar situation was encountered at LEP, where the measurement of total rate of ${\rm Z} \to \tau^+\tau^-$ events and of the tau polarization was sufficient to determine the tau couplings to the Z, regardless of initial state polarization~\cite{Altarelli:1989hv,Schael_2006}.

This letter is organized as follows. First, the reader is briefly reminded of the theoretical framework. Next, the statistical analysis of the optimal observables is described, and realistic estimates for the top-quark electroweak coupling sensitivities are obtained as a function of the centre-of-mass energy at the FCC-ee. Finally, the results are discussed and prospects for further improvements are given.

\section{Theoretical framework}
\label{sec:theory}

The top-quark couplings to the photon and the Z can be parameterized in several ways. In Ref.~\cite{Baer_2013}, for example, the analysis makes use of the usual form factors denoted $F_1$, $F_2$, defined in the following expression (with $X = \gamma, Z$):
\begin{equation}
\Gamma_\mu^{ttX} = -ie \left\{ \gamma_\mu \left( F_{1V}^X +\gamma_5 F_{1A}^X \right) + {\sigma_{\mu\nu} \over 2 m_{\rm t}} (p_t + p_{\bar t})^\nu \left( i F_{2V}^X + \gamma_5 F_{2A}^X \right)\right\},
\end{equation}
with, in the standard model, vanishing $F_2$s and
\begin{eqnarray}
F_{1V}^{\gamma} = -{2\over3} \ & , & F_{1V}^Z = {1\over 4\sin\theta_W\cos\theta_W} \left(1-{8\over3}\sin^2\theta_W\right) \ , \\ F_{1A}^{\gamma} = 0 \ & , & F_{1A}^Z = {1 \over 4\sin\theta_W\cos\theta_W} \ .
\end{eqnarray}
The sensitivities are expressed therein in terms of $\tilde{F}_1$, $\tilde{F}_2$ defined as
\begin{equation}
\tilde{F}_{1V}^X = -({F}_{1V}^X+{F}_{2V}^X)\ , \ \tilde{F}_{2V}^X = {F}_{2V}^X\ , \ \tilde{F}_{1A}^X = -{F}_{1A}^X\ , \ \tilde{F}_{2A}^X = -i{F}_{2A}^X\ .
\end{equation}

On the other hand, the optimal-observable statistical analysis presented in the next section, based on Ref.~\cite{Grzadkowski_2000}, uses the following $A,B,C,D$ parameterization (with $v = \gamma, Z$): 

\begin{equation}
\Gamma^\mu_{ttv} = {g\over 2} \left[ \gamma^\mu \left\{ (A_v+\delta A_v) - \gamma_5 (B_v+\delta B_v) \right\} + {(p_t -p_{\bar t})^\mu\over 2 m_{\rm t}}  \left( \delta C_v - \delta D_v \gamma_5 \right) \right],
\end{equation}
which easily relates to the previous parameterization with
\begin{eqnarray}
A_v+\delta A_v = - 2i\sin\theta_W \left( F_{1V}^X + F_{2V}^X \right) & \ ,  & B_v+\delta B_v = - 2i\sin\theta_W F_{1A}^X \ , \\  \delta C_v = -2i\sin\theta_W F_{2V}^X & \ , & \delta D_v = -2\sin\theta_W F_{2A}^X \ . 
\end{eqnarray}
The expected sensitivities on the anomalous top-quark couplings can be derived in any of these parameterizations. Although originally derived with that of Ref.~\cite{Grzadkowski_2000}, the final estimates presented in this study, however, use the parameterization of Ref.~\cite{Baer_2013}, for an easy comparison. For the same reason, the same restrictions as in Ref.~\cite{Baer_2013} are applied here: only the six CP conserving form factors are considered (i.e., the two $F_{2A}^X$ are both assumed to vanish), and either the four form factors $F_{1V,A}^X$ are varied simultaneously while the two $F_{2V}^X$ are fixed to their standard model values, or vice-versa. A careful reading of Ref.~\cite{Baer_2013} shows that the form factor $F_{1A}^\gamma$ was also kept to its standard model value, as a non-zero value would lead to gauge-invariance violation. It is straightforward to show that, under these restrictions, the three parameterizations lead to the same sensitivities on $F_i$, $\tilde{F}_i$ and $A,B,C,D$ (with a multiplicative factor $2\sin\theta_W \sim 0.96$ for the latter set). 

\begin{figure}[htbp]
\begin{center}
\includegraphics[width=0.8\columnwidth]{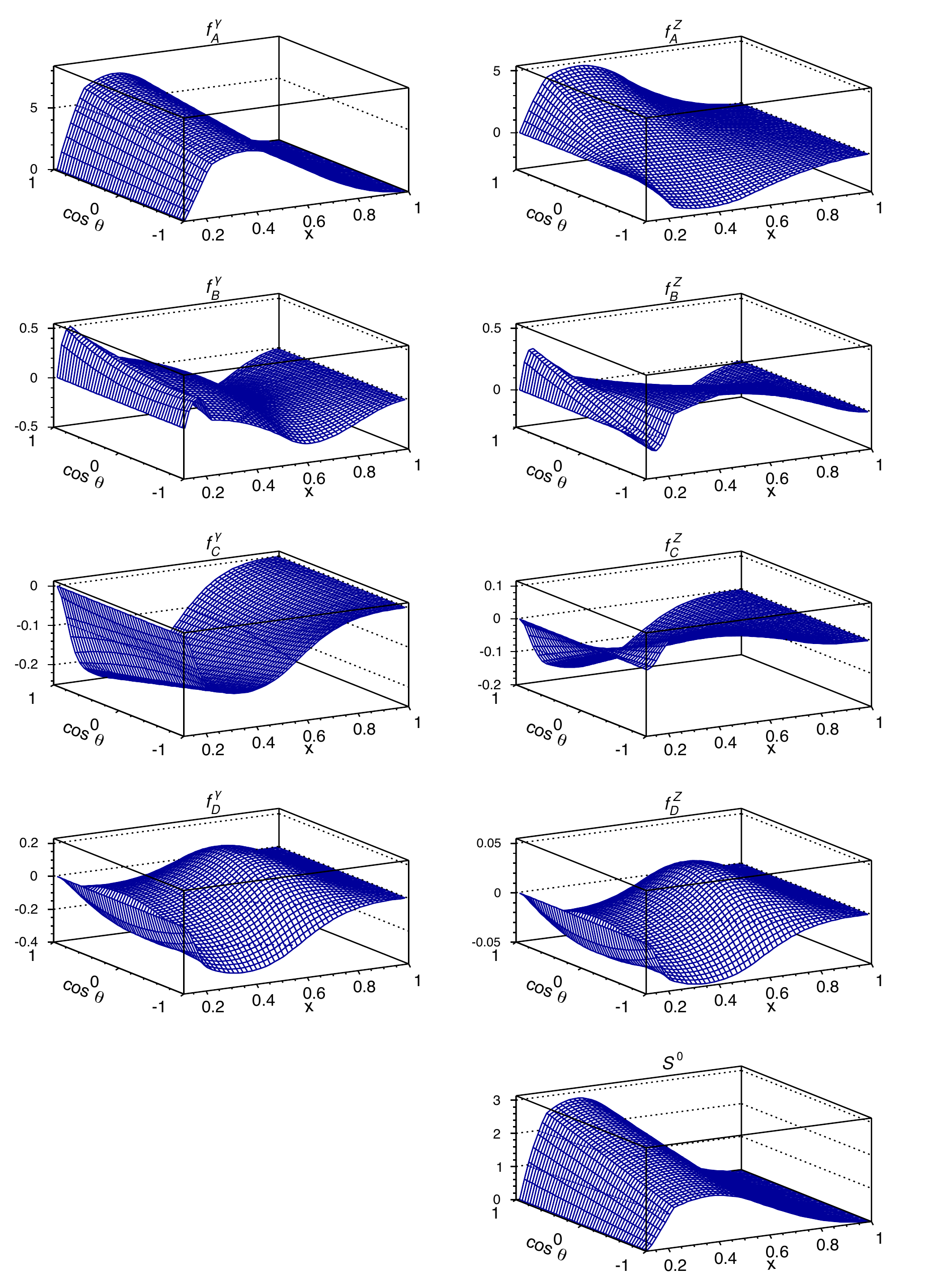}
\caption{\label{fig:distributions}
The eight $f_i(x,\cos\theta)$ functions and the standard-model contribution $S^0(x,\cos\theta)$ for $\ell^-$ at $\sqrt{s} = 365$\,GeV. Left column, from top to bottom: $f_1 = f_A^\gamma$~; $f_3 = f_B^\gamma$~; $f_5 = f_C^\gamma$~; and $f_7 = f_D^\gamma$. Right column, from top to bottom: $f_2 = f_A^Z$~; $f_4 = f_B^Z$~; $f_6 = f_C^Z$~; $f_8 = f_D^Z$~; and $S^0$. In all these figures, $\theta$ is the lepton polar angle, and $x$ is the reduced lepton energy, defined as $x = {2 E_\ell \over m_t} \sqrt{1-\beta \over 1+ \beta}$, where $\beta$ is the top velocity and $m_t$ is the top mass.}
\end{center}
\end{figure}

The tree-level angular and energy distributions of the lepton arising from the ${\rm t \bar t}$ semi-leptonic decays are known analytically as a function of the incoming beam polarizations and the centre-of-mass energy~\cite{Grzadkowski_2000}: 
\begin{equation}
{{\rm d}^2\sigma \over {\rm d}x {\rm d}\cos\theta} = {3\pi\beta\alpha^2(s) \over 2s} B_\ell S_\ell(x,\cos\theta),
\end{equation}
where $\beta$ is the top velocity, $s$ is the centre-of-mass energy squared, $\alpha(s)$ is the QED running coupling constant, and $B_\ell$ is the fraction of ${\rm t\bar t}$ events with at least one top quark decaying to either ${\rm e}\nu_{\rm e}{\rm b}$ or $\mu\nu_\mu{\rm b}$ (about 44\%). As the non-standard form factors $\delta(A,B,C,D)_v \equiv \delta_i$ are supposedly small, only the terms linear in $\delta_i$ are kept: 
\begin{equation}
\label{eq:optimal}
S(x,\theta) = S^0(x,\theta) + \sum_{i=1}^8 \delta_i f_i(x,\cos\theta),
\end{equation}
where $x$ and $\theta$ are the lepton (reduced) energy and polar angle, respectively, and $S^0$ is the standard-model contribution. The eight distributions $f_{A,B,C,D}^{\gamma, Z}(x,\cos\theta) \equiv f_i(x,\cos\theta)$ and the standard-model contribution $S^0(x,\cos\theta)$ are shown for $\ell^-$ in Fig.~\ref{fig:distributions} at $\sqrt{s} = 365$\,GeV, with no incoming beam polarization. 

\section{Optimal-observable statistical analysis}
\label{sec:optimal}

There are nine different functions entering Eq.~\ref{eq:optimal}, and eight form factors $\delta_i$ to be evaluated from a given sample of ${\rm t \bar t}$ events. In principle, all eight form factors and their uncertainties can therefore be determined simultaneously, under the condition that the nine functions are linearly independent. Experimentalists usually maximize numerically a global likelihood $L$ -- or equivalently, minimize the negative Log-likelihood ($-\log L$) -- with respect to all form factors:
\begin{equation}
\label{eq:likelihood}
L = {\mu^N \over N!}{\rm e}^{-\mu} \times \prod_{k=1}^N p(k),
\end{equation} 
where $N$ is the total number of ${\rm t \bar t}$ events observed in the data sample, $\mu$ is the number of events expected for the integrated luminosity ${\cal L}$ of the data sample ($\mu = \sigma_{\rm tot} \times {\cal L}$), and
\begin{equation} 
p(k)  = {1 \over \sigma_{\rm tot} } { {\rm d}^2\sigma \over {\rm d}x {\rm d}\cos\theta }(x_k,\cos\theta_k) {\rm , with \ } 
\sigma_{\rm tot} =  \int { {\rm d}^2\sigma \over {\rm d}x {\rm d}\cos\theta } {\rm d}x {\rm d}\cos\theta .
\end{equation}
The covariance matrix obtained from the numerical minimization of the negative log-likelihood is then inverted to get the uncertainties on the form factors, $\sigma(\delta_i)$. It can be shown~\cite{Davier_1993} that, in the linear form given in Eq.~\ref{eq:optimal}, this method is statistically optimal for the determination of the $\sigma(\delta_i)$.  The functions $f_i(x,\cos\theta)$ are therefore called "optimal observables". It turns out~\cite{Diehl_1994} that the covariance matrix, hence the statistical uncertainties on the form factors, can be obtained analytically in the limit of a large number of events, which is the case considered in this letter. Specifically, if the total event rate is included in the derivation of the likelihood as is the case in Eq.~\ref{eq:likelihood}, the elements of the covariance matrix $V$ are given by (${\rm d}\Omega \equiv {\rm d}x {\rm d}\cos\theta$)
\begin{equation}
\label{eq:rate}
V_{ij} = {\cal L} \int {\rm d}\Omega { f_i \times f_j \over S^0}  \ ,
\end{equation}
while if the total event rate is not included in the likelihood, namely by removing the first term of the product in Eq.~\ref{eq:likelihood}, these elements take the form
\begin{equation}
\label{eq:norate}
V_{ij} = {\cal L} \left[ \int {\rm d}\Omega {f_i \times f_j \over S^0}  -  { \int {\rm d}\Omega  f_i  \int {\rm d}\Omega  f_j \over \int {\rm d}\Omega S_0 } \right],
\end{equation}
and the uncertainty on the form factor $\delta_i$ is simply 
\begin{equation}
\sigma(\delta_i) = \sqrt{ [ V^{-1} ]_{ii} } \ .
\end{equation}
This analytical procedure is used in Ref.~\cite{Grzadkowski_2000} to determine the sensitivity to top-quark electroweak couplings in $500\,{\rm fb}^{-1}$ of ${\rm e^+ e^-}$ collisions at $\sqrt{s} = 500$\,GeV, with or without incoming beam polarization. In this article, the authors evaluate the covariance matrix with Eq.~\ref{eq:rate}, but they let the total number of events float by adding a fictitious multiplicative form factor $\delta_0$ in front of $S^0$ in Eq.~\ref{eq:optimal}, hence increase the rank of the covariance matrix from 8 to 9. It was checked that this work-around is numerically equivalent to using Eq.~\ref{eq:norate}, {\it i.e.}, to not use the total event rate in the likelihood. 

A quick survey of Fig.~\ref{fig:distributions}, however, shows that $f_A^\gamma(x,\cos\theta)$, in the top-left corner, is almost degenerate with the standard model contribution $S^0(x,\cos\theta)$, in the bottom-right corner. Letting the normalization of the standard model contribution float is therefore bound to lead to very large statistical uncertainties on all form factors, as is indeed observed in Ref.~\cite{Grzadkowski_2000}. For this reason, and as is done in Ref.~\cite{Baer_2013}, the present study includes the total event rate in the determination of the uncertainties.

As already mentioned, it is possible to determine simultaneously all eight form factors and their uncertainties. In the first configuration of Ref.~\cite{Baer_2013}, however, only the three coefficients $F_{1V}^\gamma$, $F_{1V}^Z$ and $F_{1A}^Z$ are allowed to vary. The other five form factors are fixed to their standard model values. In this simplified situation, Eq.~\ref{eq:optimal} reads
\begin{equation}
S(x,\theta) = S^0(x,\theta) - 2i\sin\theta_W \delta F_{1V}^\gamma  f_A^\gamma - 2i\sin\theta_W \delta F_{1V}^Z f_A^Z +  -2i\sin\theta_W \delta F_{1A}^Z f_B^Z \ ,
\end{equation}
which leads to the following $3\times 3$ covariance matrix $V_1 = 4\sin^2\theta_W \times {\cal L} \times X$, with
\begin{eqnarray}
\label{eq:firstInt} X_{11} = \int {\rm d}\Omega {(f_A^\gamma)^2 \over S^0} \ , \ & {\displaystyle X_{12} = \int {\rm d}\Omega {f_A^\gamma \times f_A^Z \over S^0}} \ , \ & X_{13} = \int {\rm d}\Omega {f_A^\gamma \times f_B^Z \over S^0} \ , \\
& {\displaystyle X_{22} = \int {\rm d}\Omega {(f_A^Z)^2 \over S^0} } \ , \ & X_{23} = \int {\rm d}\Omega {f_A^Z \times f_B^Z \over S^0} \ , \\
& & X_{33} = \int {\rm d}\Omega {(f_B^Z)^2 \over S^0} \ .
\end{eqnarray}
In the second configuration of Ref.~\cite{Baer_2013}, only the two coefficients $F_{2V}^\gamma$ and $F_{2V}^Z$ are allowed to vary, which leads to the even simpler expression of Eq.~\ref{eq:optimal}:
\begin{equation}
S(x,\theta) = S^0(x,\theta) - 2i\sin\theta_W \delta F_{2V}^\gamma  (f_A^\gamma + f_C^\gamma) - 2i\sin\theta_W \delta F_{2V}^Z  (f_A^Z + f_C^Z)\ ,
\end{equation}
and the following $2\times 2$ covariance matrix $V_2 = 4\sin^2\theta_W \times {\cal L} \times Y$, with 
\begin{eqnarray}
Y_{11} = \int {\rm d}\Omega {(f_A^\gamma + f_C^\gamma)^2 \over S^0} \ , \ & {\displaystyle Y_{12} = \int {\rm d}\Omega {(f_A^\gamma + f_C^\gamma) \times (f_A^Z + f_C^Z) \over S^0}} &  \ , \\
\label{eq:lastInt} & {\displaystyle Y_{22} = \int {\rm d}\Omega {(f_A^Z + f_C^Z)^2 \over S^0} } & \ .
\end{eqnarray}
The numerical results are presented in the next section for the case of the FCC-ee. 

\section{Sensitivity to the top-quark electroweak couplings}
\label{sec:sensitivities}

The aforementioned covariance matrices assume a perfect event reconstruction, an event selection efficiency of 100\%, a $4\pi$ detector acceptance, and the absence of background processes. While these hypotheses would not be utterly unrealistic at an ${\rm e^+ e^-}$ collider, a more conservative approach is in order to render the present estimates credible and reliable. 

{\bf Event reconstruction} 

The only reconstructed quantities used for the determination of the covariance matrices are the lepton direction and the lepton energy (or momentum). Both quantities can be reconstructed with more than adequate precision, as was the case with the detectors built for the LEP collider. The numerical evaluation of the integrals in Eqs.~\ref{eq:firstInt} to~\ref{eq:lastInt} are however performed with 50 bins in $x$ and $\cos\theta$. This procedure corresponds to conservatively assuming a lepton energy resolution of 1\,GeV and a lepton angular resolution of 20\,mrad, figures vastly exceeded by LEP detectors.

{\bf Event selection and particle identification}

The event selection relies on the presence of an energetic isolated lepton and two energetic b-quark jets in the final state, accompanied by either two light-quark jets or an additional lepton. At $\sqrt{s} = 365$\,GeV, the lepton momentum can take values between 13.5 and 120\,GeV/$c$, a range in which an identification efficiency of 80\% can be conservatively assumed, with a negligible fake rate.  Similarly, the b-quark jet energies can take values between 49 and 94\,GeV, for which b-tagging algorithms are both efficient and pure, especially with two b jets in the final state. A very conservative b-tagging efficiency of 60\% is assumed here. To emulate these efficiencies, all terms of Eq.~\ref{eq:optimal}, hence all covariance matrix elements, are multiplied by $0.6 \times 0.8 = 0.48$. 

{\bf Detector acceptance}

The polar-angle coverage of a typical detector at ${\rm e^+ e^-}$ colliders is usually assumed to be from 10 to 170 degrees. To be conservative, the leptons are assumed here to be detected only for $ | \cos\theta |< 0.9$, {\it i.e.}, in a range from 26 to 154 degrees. This effect is emulated by evaluating the integrals of Eqs.~\ref{eq:firstInt} to~\ref{eq:lastInt} between $\cos\theta_{\rm min} = -0.9$ and  $\cos\theta_{\rm max} = 0.9$. Given the large value of the minimum lepton energy, the integration bounds over $x$ are left untouched. 

{\bf Background processes}

The major background identified in Ref.~\cite{Amjad_2013} (which Ref.~\cite{Baer_2013} is based upon) is the single-top production in association with a W boson and a b quark, through WW$^\ast$ production, as it leads to the same final state as the top-quark pair production. The corresponding cross section~\cite{Boos_2001} increases fast with the centre-of-mass energy, and critically depends on the incoming beam polarization. At $\sqrt{s} = 500$\,GeV, the single-top production cross section can reach up to 20\% of the top-pair production cross section in the final state with an electron or a positron and in the ${\rm e^-_L e^+_R}$ initial polarization configuration. Yet, this background has not been included in the top-quark electroweak coupling study of Ref.~\cite{Baer_2013}. At $\sqrt{s} = 365$\,GeV and with unpolarized beams, however, the single-top cross section in the same final state amounts to about 0.1\% of the pair production cross section. It was therefore ignored for the first estimate of precisions given below.. 

{\bf Other experimental uncertainties}

A number of other experimental uncertainties are listed in Ref.~\cite{Amjad_2013}, such as those affecting the measurement of the beam polarization (which enters crucially the cross section measurement); the effects of beamstrahlung; or the ambiguous top-quark reconstruction (which enters crucially the forward-backward asymmetry measurement). These uncertainties apply neither to the FCC-ee, where beamstrahlung effects are negligible and no beam polarization needs to be measured, nor to the present study, as the top-quark direction needs not be reconstructed. The experimental uncertainties affecting the lepton energy and angular distributions can be safely neglected, given the conservative assumptions on the resolutions. The total event rate, needed for the present study, requires a precise luminosity determination, a measurement that can be controlled to a fraction of a per mil, hence neglected here. 

{\bf Theoretical uncertainties}

The dominant systematic uncertainty is of theoretical nature. The total event rate indeed requires an accurate prediction of the total cross section for top pair production. The precision of this prediction is inferred to be at the level of a few per mil in Ref.~\cite{Amjad_2013} for $\sqrt{s} = 500$\,GeV. A similar precision can be expected at smaller centre-of-mass energy as long as it is reasonably above the production threshold.  

{\bf Integrated luminosity profile}

\begin{figure}[htbp]
\begin{center}
\includegraphics[width=0.8\columnwidth]{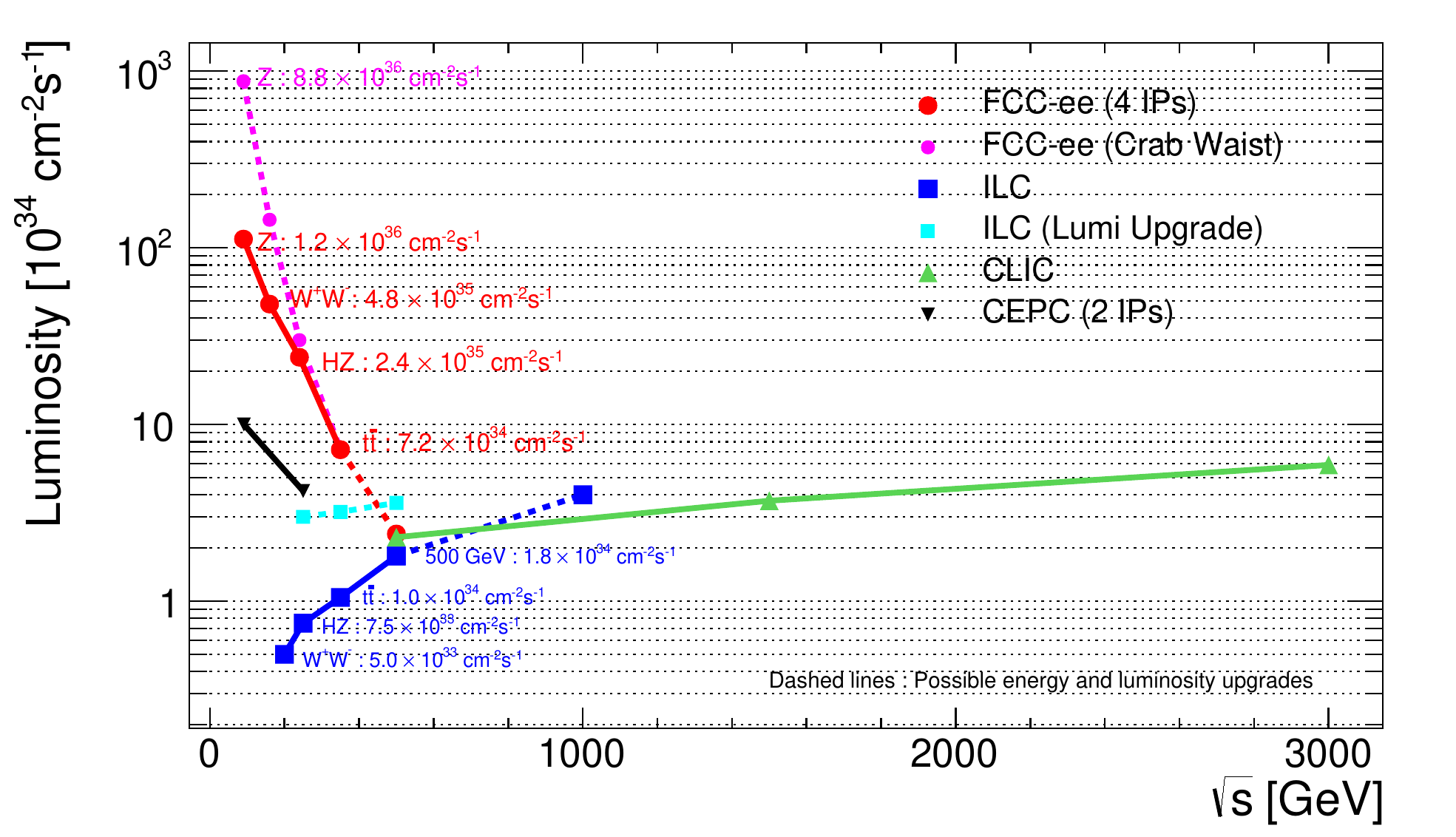}
\caption{\label{fig:lumi}
The target luminosities at the FCC-ee, as a function of the centre-of-mass energy: red (baseline beam crossing) and purple (crabbed-waist beam crossing) lines. The plot also indicates the target luminosities of of other ${\rm e^+ e^-}$ colliders under study in the world. Figure taken from the FCC-ee official web site~\cite{FCCeeWebSite}.}
\end{center}
\end{figure}

The target luminosities at the FCC-ee are displayed in Fig.~\ref{fig:lumi}~\cite{FCCeeWebSite} as a function of the centre-of-mass energy, together with the target luminosities of other ${\rm e^+ e^-}$ colliders under study in the world. At $\sqrt{s} = 350$\, GeV, a luminosity of $7.2 \times 10^{34}\, {\rm cm}^{-2}{\rm s}^{-1}$ is expected to be democratically distributed to four interaction regions, leading to an integrated luminosity of $3.6\, {\rm ab}^{-1}$ over a period of five years. About $1{\rm ab}^{-1}$ ought to be kept for threshold measurements (leading to a statistical precision on the top mass of about 15\,MeV), and the rest can be used to perform measurements above the production threshold. 

The maximum centre-of-mass energy of the FCC-ee is yet unknown. It was inferred in Ref.~\cite{Bicer_2014} that, if the total RF voltage were increased by a factor 3 with respect to the baseline, a centre-of-mass energy of 500\,GeV could be reached, and an integrated luminosity of $500\, {\rm fb}^{-1}$ could be delivered over a period of three years, as displayed in Fig.~\ref{fig:lumi} with the red dashed line. In the framework of the FCC, however, the interest of such an upgrade could not be demonstrated for the physics of the Higgs boson~\cite{Bicer_2014}. It is interesting to re-evaluate this statement in view of the physics of the top quark.

\begin{figure}[htbp]
\begin{center}
\includegraphics[width=0.85\columnwidth]{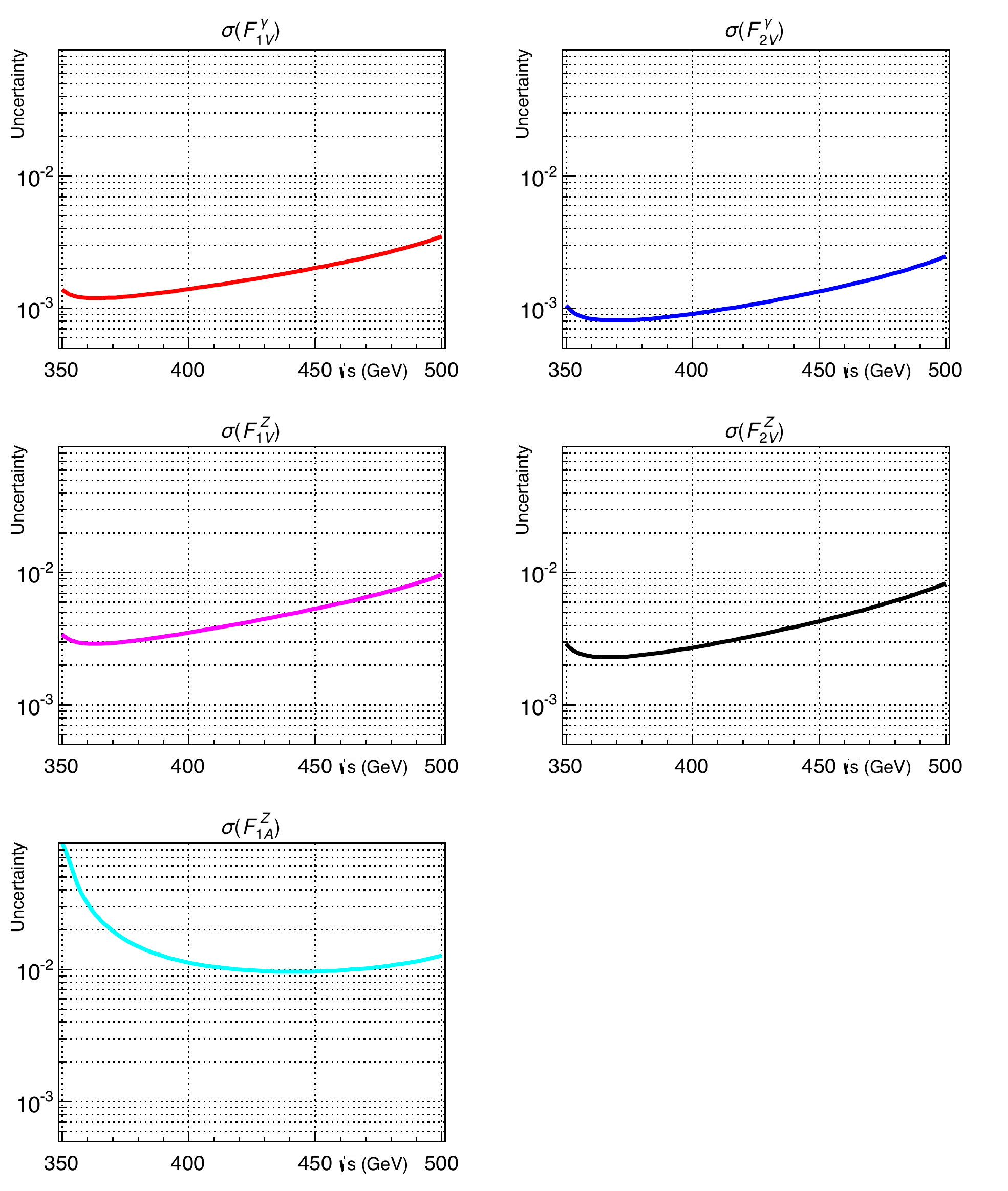}
\caption{\label{fig:variation}
Variation with the centre-of-mass energy of the statistical uncertainties of the five top-quark electroweak form factors considered in Ref.~\cite{Baer_2013}, at the FCC-ee. Left column, from top to bottom: $F_{1V}^\gamma$, $F_{1V}^Z$, and $F_{1A}^Z$. Right column: $F_{2V}^\gamma$ and $F_{2V}^Z$.}
\end{center}
\end{figure}

The centre-of-mass energy was therefore varied from 350 to 500\,GeV, and the corresponding integrated luminosity was varied linearly with $\sqrt{s}$ from 2.6 to 0.5\,${\rm ab}^{-1}$. The expected uncertainties on the top electroweak form factors, $\sigma(F_{1V}^\gamma)$, $\sigma(F_{1V}^Z)$, $\sigma(F_{1A}^Z)$, $\sigma(F_{2V}^\gamma)$ and $\sigma(F_{2V}^Z)$, were determined as explained in the previous section, with corrections for the lepton energy and angular resolutions, the event selection efficiency, and the detector acceptance,  as described above, for each value of the centre-of-mass energy. The variation of these uncertainties with $\sqrt{s}$ is shown in Fig.~\ref{fig:variation}. 

The first striking observation is that an increase of the centre-of-mass energy far beyond the top-pair production threshold is not particularly relevant to improve the precision on the top-quark electroweak couplings, as already pointed out in Ref.~\cite{Vos_2015}. For four out of five couplings, optimum precision is actually reached for $\sqrt{s} \simeq 365$\,GeV, and for the fifth one the precision is within 50\% of optimum at this energy. The expected precision then degrades by up to a factor four with 500\,${\rm fb}^{-1}$ at $\sqrt{s} = 500$\,GeV. It can also be noted that a very decent precision is alredy reached for $\sqrt{s} = 350$\,GeV. The second observation is that the precision reached for these four couplings is at the level of the per mil, and that the tt$\gamma$ and the ttZ couplings can be determined independently with this precision without the need of initial polarization. 

It is only for $F_{1A}^Z$ that a moderate increase of the centre-of-mass energy would improve the precision by a factor of two, from 2.2\% at $\sqrt{s} = 365$\,GeV to 1\% at $\sqrt{s} = 440$\,GeV, an energy at which the single-top production would need to be included as a background to the study. There are, however, many other observables to be studied in a ${\rm t\bar t}$ event, beyond the energies and angles of the leptons. It was noticed, for example, that a factor of two improvement could be obtained for $\sigma(F_{1A}^Z)$ at $\sqrt{s} = 365$\,GeV with the energy and angular distributions of the b quarks instead of the leptons. The use of the b jets will be the subject of further studies with more detailed event reconstruction algorithms.

\section{Results and discussion}
\label{sec:results}

{\bf Expected statistical accuracies}

\begin{figure}[htbp]
\begin{center}
\includegraphics[width=1\columnwidth]{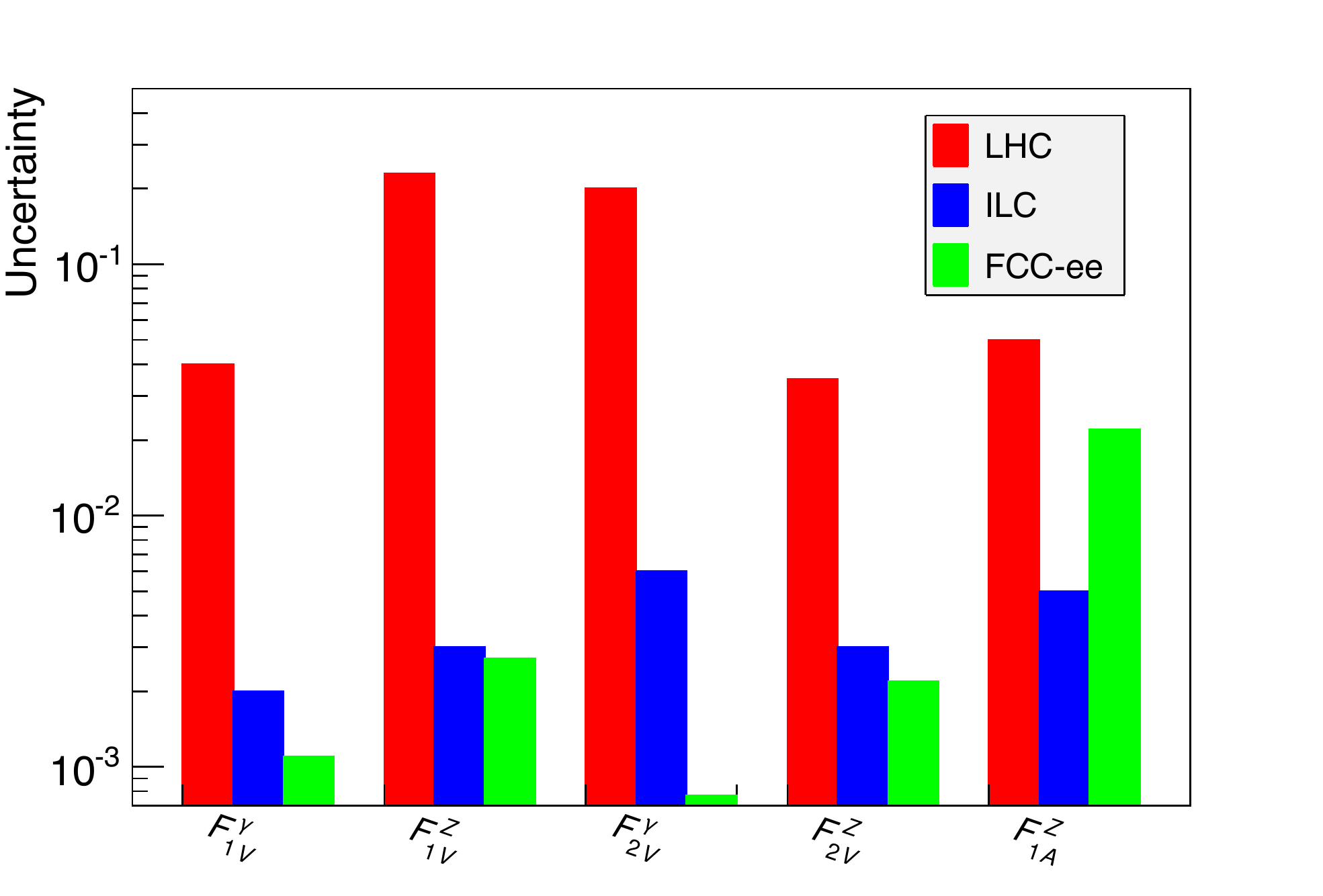}
\caption{\label{fig:baer}
(Modified from Ref.~\cite{Baer_2013}). Statistical uncertainties on CP-conserving top-quark form factors expected at the ILC (blue) and the LHC (red). The figure was modified to include the projections from the FCC-ee. The results for the LHC assume an integrated luminosity of 300\,${\rm fb}^{-1}$ and a centre-of-mass energy of 14\,TeV. The results for the ILC assume an integrated luminosity of  500\,${\rm fb}^{-1}$ at $\sqrt{s} = 500$\,GeV, and beam polarizations of ${\cal P} = \pm 0.8$, ${\cal P}^\prime = \mp 0.3$. The ILC projections are obtained from the measurements of the total top-quark pair production cross section, together with the top-quark forward-backward asymmetry. The FCC-ee projections are obtained at $\sqrt{s}=365$\,GeV, with unpolarized beams and with an integrated luminosity of 2.4\,${\rm ab}^{-1}$, from the sole lepton angular and energy distributions.}
\end{center}
\end{figure}

A picture is often better than many words. This study is best summarized by Fig.~\ref{fig:baer}, taken from Ref.~\cite{Baer_2013}, and modified by the addition of the FCC-ee projections at $\sqrt{s} = 365$\,GeV. As anticipated, the lack of incoming beam polarization at the FCC-ee is more than compensated by the use of the final state polarization and by a significantly larger integrated luminosity, even with the sole use of the lepton energy and angular distributions, and modest detector performance.

{\bf Theory uncertainties}

\begin{figure}[htbp]
\begin{center}
\includegraphics[width=0.95\columnwidth]{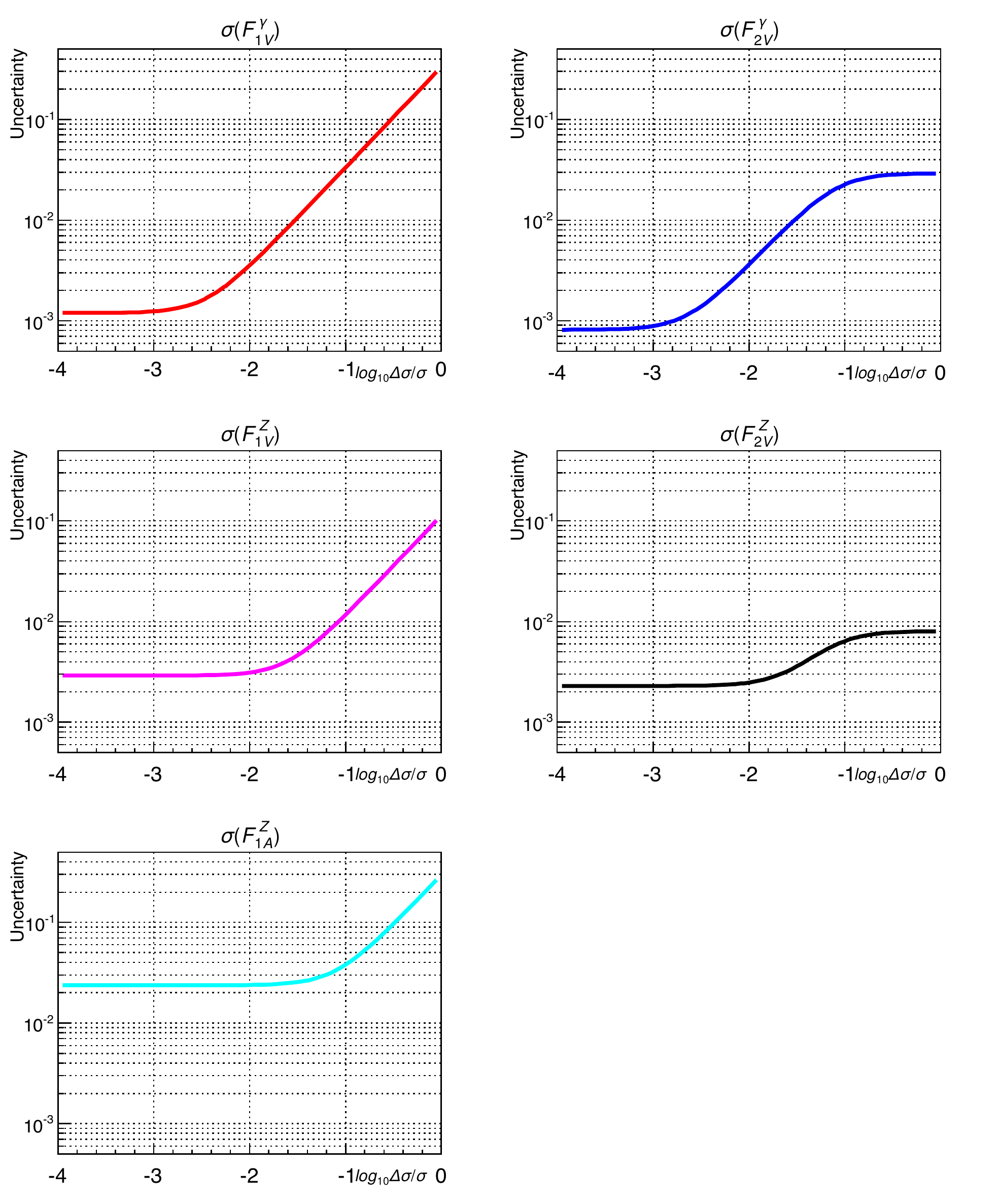}
\caption{\label{fig:crosserr}
Uncertainty on the form factors at the FCC-ee with 2.4\,${\rm ab}^{-1}$ at $\sqrt{s} = 365$\,GeV, as a function of the relative cross-section theorerical error, varied from 0.01\% to 100\%. Left column, from top to bottom: $F_{1V}^\gamma$, $F_{1V}^Z$, and $F_{1A}^Z$. Right column: $F_{2V}^\gamma$ and $F_{2V}^Z$.}
\end{center}
\end{figure}

As mentioned in the previous section, the dominant systematic error on these numbers is the theoretical uncertainty on the predicted event rate. It is difficult to say today what this uncertainty will be at the time of the FCC-ee startup. To evaluate its effects, the likelihood in Eq.~\ref{eq:likelihood} was enhanced with the corresponding Gaussian nuisance factor, and the form factor uncertainties were determined for any value of the assumed cross-section theoretical error. The result is displayed in Fig.~\ref{fig:crosserr} for a theoretical error between 0.01\% and 100\%. The uncertainties on the first four form factors stay below a few per mil if the total cross section can be predicted with a precision of 2\% or better. The uncertainty on $F_{1A}^Z$ remains essentially unaffected as long as the theoretical precision on the cross section is below 10\%.

{\bf Discussion}

The above results are obtained under the assumption that the gauge-invariance-violating form factor ($F_{1A}^\gamma$) and the CP-violating form factors ($F_{2A}^{\gamma,Z}$) vanish, to allow for a one-to-one and straightforward comparison with Ref.~\cite{Baer_2013}. From an experimental point-of-view, however, there is no a-priori reason why these form factors could not be extracted from the measurements of the lepton angular and energy distributions. The present study is therefore extended, with $2.4\,{\rm ab}^{-1}$ at $\sqrt{s} = 365$\,GeV, to the following two configurations by relaxing the constraints on $F_{1A}^\gamma$, $F_{2A}^\gamma$ and $F_{2A}^Z$: either the four form factors $F_{1V,A}^X$ are varied simultaneously while the four $F_{2V,A}^X$ are fixed to their standard model values, or vice-versa. 

In the first configuration, it turns out that relaxing the constraint on $F^\gamma_{1A}$ does not sizeably change the precision on the other three $F^X_{1V,A}$ form factors, as shown in Table~\ref{tab:f1}. A per-cent accuracy is also obtained on $F^\gamma_{1A}$.
\begin{table}[htbp]
\begin{center}
\caption{\label{tab:f1} Precision on the four $F_{1V,A}^X$ expected with $2.4\,{\rm ab}^{-1}$ at $\sqrt{s} = 365$\,GeV at the FCC-ee, if $F_{1A}^\gamma$ is fixed to its standard model value (first row) or if this constraint is relaxed (second raw). The precisions expected with $500\,{\rm fb}^{-1}$ at $\sqrt{s} = 500$\,GeV are indicated in the third row.}
\begin{tabular}{|l|l|l|l|l|}
\hline Precision on & $F_{1V}^\gamma$ & $F_{1V}^Z$ & $F_{1A}^\gamma$ & $F_{1A}^Z$ \\
\hline\hline Only three $F_{1V,A}^X$ &     $1.2\, 10^{-3}$     &    $2.9\, 10^{-3}$     &   $0.0\, 10^{-2}$    &   $2.2\, 10^{-2}$ \\
\hline All four $F_{1V,A}^X$  & $1.2\, 10^{-3}$ & $3.0\, 10^{-3}$ & $1.3\, 10^{-2}$ & $2.4\, 10^{-2}$ \\ 
\hline $\sqrt{s} = 500$\,GeV  & $5.5\, 10^{-3}$ & $1.5\, 10^{-2}$ & $1.0\, 10^{-2}$ & $2.2\, 10^{-2}$ \\ 
\hline
\end{tabular}
\end{center}
\end{table}

The situation with the $F_{2V,A}^X$ form factors in the second configuration is even clearer. Indeed, the distributions related to $F_{2A}^\gamma$ and $F_{2A}^Z$ form factors are CP odd, while those related to $F_{2V}^\gamma$ and $F_{2V}^Z$ are CP even. With vanishing correlation coefficients, the two pairs of form factors can therefore be determined independently from each other. The precisions on $F_{2V}^\gamma$ and $F_{2V}^Z$, expected with $2.4\,{\rm fb}^{-1}$ at $\sqrt{s} = 365$\,GeV at the FCC-ee, are thus unchanged with respect to Fig.~\ref{fig:baer} when the constraint on $F_{2A}^\gamma$ and $F_{2A}^Z$ is relaxed, and amount to $8.1\, 10^{-4}$ and $2.3\, 10^{-3}$ respectively. With $500\,{\rm fb}^{-1}$ at $\sqrt{s} = 500$\,GeV, the precisions would be $2.5\, 10^{-3}$ and $8.3\, 10^{-3}$ respectively, as also shown in Fig.~\ref{fig:variation}.

The accuracy of the CP-violating form factors with the sole lepton angle and energy distributions is moderately constraining ($1.4\, 10^{-1}$ on $F_{2A}^\gamma$ and $8.2\, 10^{-1}$ on $F_{2A}^Z$) because of the important correlation between the two distributions $f_D^\gamma$ and $f_D^Z$, well visible in Fig.~\ref{fig:distributions}. A relevant precision of $1.7\, 10^{-2}$, however, is reached on the linear combination $F_{2A}^\gamma + 0.17 \times F_{2A}^Z$ with $2.4\,{\rm ab}^{-1}$ at $\sqrt{s} = 365$\,GeV, reduced to $0.9\, 10^{-2}$ with $500\,{\rm fb}^{-1}$ at $\sqrt{s} = 500$\,GeV. A reduction of the correlation between these two form factors requires the analysis of additional observables, beyond the scope of the present study.

Similarly, when all eight parameters are considered simultaneously, the lepton angle and energy distributions are no longer sufficient to avoid large correlations between form factors. The same observation was made in Refs.~\cite{Baer_2013} and~\cite{Amjad_2013} with the four observables chosen for the analysis at 500\,GeV and with incoming beam polarizations. A generator-level exercise with more observables in the fully leptonic final state has been recently attempted in Ref.~\cite{ledib_2015}, released after the present study. In this exercise, an optimal-observable analysis of the matrix element squared is carried out with thirteen different observables (the top quark direction, the $\ell^+$ and $\ell^-$ angles and energies, the ${\rm b}$ and ${\rm \bar b}$ angles and energies, and the invariant masses of the top quarks and W bosons), with unambiguous identification and reconstruction under the assumption of a perfect detector. With these additional variables, the few degeneracies between form factors are indeed removed, but the conclusion is identical to that of this paper: the incoming beam polarizations are not essential in the process. 

A similar analysis could be undertaken for semi-leptonic final states at $\sqrt{s} = 365$\,GeV, in order to determine all eight form factors simultaneously with the ultimate accuracy, but the assumption of a perfect detector cannot be expected to give fully reliable results when the jets and the missing energy from the top decays are to be included, as acknowledged in Ref.~\cite{ledib_2015}. Such an analysis will be carried out when a complete simulation and reconstruction in a realistic detector becomes available for the FCC-ee study.  

\section{Summary and outlook}
\label{sec:summary}

In this paper, it has been shown that the measurements of the angular and energy distributions in semi-leptonic ${\rm t\bar t}$ events (${\rm e^+e^- \to t \bar t} \to \ell \nu {\rm q \bar q b \bar b}$) at future ${\rm e^+e^-}$ colliders have a strong potential for a precise determination of the top-quark electroweak couplings.  It has been demonstrated, even with the sole use of these two distributions and modest detector performance, that the lack of incoming beam polarization at the FCC-ee is compensated by the polarization of the final state top quarks, and by a significantly larger integrated luminosity.

Although these projections were obtained with somewhat conservative hypotheses on the detector performance, it will also be important to reproduce the results with a full simulation in a Monte Carlo study, as to further investigate that the detector requirements are indeed quite modest. While the inferred precisions are already competitive with other projects on the market, such a Monte Carlo study will also allow a reliable reconstruction of all observables in the event, beyond the lepton energies and directions, and is expected to bring sizeable improvements, especially on the few remaining correlations between form factors.

The present study is only a first look at this topic for the FCC-ee. It enhances the fantastic potential of a 100-km circular ${\rm e^+e^-}$ collider already envisioned in Ref.~\cite{Bicer_2014} with the full profiling of the top quark from a precise measurement of its electroweak couplings. In view of these new estimates, it becomes of particular interest to check their added value to the sensitivity to new physics, especially when combined with the unequalled precision of the measurements of the Z, the W, and the Higgs boson properties, as well as of the top-quark mass, at the FCC-ee.  

\section*{Acknowledgments}
I would like to give credit where credit is due: this work would not have been undertaken without the scientific vision of Alain Blondel, and his firm belief that the final state polarization would suffice to disentangle the tt$\gamma$ and ttZ couplings. 

A great deal of inspiration also came from the passionate debates about the merits of the FCC with my former thesis advisor and scientific mentor, Fran{\c c}ois Le Diberder, and my former lab director, Fran{\c c}ois Richard. I am grateful to both. 

I would like to sincerely thank Zenro Hioki for digging out his code written 15 years ago, running it for me, and allowing me, with his patient explanations, to understand why my uncertainties were so much smaller than those obtained in Ref.~\cite{Grzadkowski_2000}. 

Finally, I am indebted to Patrizia Azzi, Alain Blondel, Christophe Grojean, and Roberto Tenchini, for their careful reading of the manuscript and for the wisdom of their comments and suggestions. 

\vfill\eject

\bibliography{converted_to_latex.bib}

\end{document}